\documentclass{article}
\usepackage{amsmath,mathrsfs}
\usepackage{graphicx,subfigure}
\usepackage{multirow,booktabs}
\usepackage[numbers,sort&compress]{natbib}
\usepackage[colorlinks,linkcolor=blue,citecolor=red,urlcolor=blue]{hyperref}

\def\abstract#1{\vskip 7mm 
        \begin{center}{\large Abstract}\par \smallskip
                \begin{minipage}[c]{12cm}
                        \small #1
                \end{minipage}
        \end{center}
}
\def\title#1{\begin{center}{\Large\bf #1}\end{center}}
\def\author#1{\vskip 5mm \begin{center}{#1}\end{center}}
\def\address#1{\begin{center}{\it #1}\end{center}}
\makeatletter
\@ifundefined{lesssim}{}{}
\@ifundefined{gtrsim}{}{}
\def\vereq#1#2{\lower3pt\vbox{\baselineskip1.5pt \lineskip1.5pt
\ialign{$\m@th#1\hfill##\hfil$\crcr#2\crcr\sim\crcr}}}
\makeatother

\begin{document}

\title{%
  Large Scale Lorentz Violation Gravity and Dark Energy
}
\author{ Jiayin Shen$^a$ and Xun Xue$^{a,b,}$\footnote{Corresponding author:xxue@phy.ecnu.edu.cn}}
\address{ $^a$Department of Physics, East China Normal University, Shanghai 200241, China}
\address{$^b$Center for Theoretical Physics, Xinjiang University, Urumqi 830046, China}

\abstract{
 The accelerating expansion of universe can be described by the non-zero cosmological constant or the dark energy. However, the origin of the dark energy remains a mystery of modern physics. The local Lorentz invariance is the most exact symmetry of the Nature on the one hand, but all quantum gravity theories predict Lorentz violation on the other hand. The local Lorentz violation induced by the quantum gravity at the very early universe may be transformed into large scale by inflation. Combining the low-$l$ anomalies of the  CMB spectrum, we propose that the local Lorentz invariance may be broken at the large scale. We construct the effective gravity at the cosmic scale with a local $SO(3)$ symmetry. The theory exhibits non-trivial contortion distribution even with scalar matter source. The FRW like solution of the theory is analyzed and the contortion distribution contributes a dark energy like effect which is responsible for the accelerating expansion of the universe. It reveals that the dark energy may be the remnants of quantum gravity in this sense.
}

\section{Introduction}

Though general relativity(GR) passes almost all observational examination, it is still controversial whether GR is the ultimate gravitational theory at all macroscopic scale even to cosmic one. There indeed observations deviating from predictions of GR at large scale. One is the galaxy rotation curve problem which can be explained with the existence of dark matter besides the luminous matter. The other is the discovery of accelerating expansion of the universe in 1998\cite{filippenko1998results,riess1998observational,perlmutter1999measurements}, which can be described by introducing a small positive cosmological constant $\Lambda$ into the Einstein field equation or adjoining "dark energy" with ${\rho\simeq-p}$ to the energy-momentum tensor of cosmic media, known as $\Lambda$CDM model\cite{hinshaw2013nine}. Although there are many dark energy models such as quintessence, phantom, etc, its origin is still a mystery of physics and there is no reasonable candidate for it. Therefore, while $\Lambda$CDM model has been a great success, dark energy's origin and essence is being the deep darkness, just as its name. On the other hand, all searches for the dark matter particles gives negative results until now.

It is reasonable to explore the deviation from GR at scale larger than the galaxy to the cosmic one with modified gravity rather than the assumption of dark matter or dark energy considering that the GR is only verified within the solar system or for the local phenomenon. Most of popular modified gravity models assume the local Lorentz invariance just as in the GR which can be regarded as a Lorentz gauge theory\cite{utiyama1956invariant,sciama1962analogy,kibble1961lorentz,trautman1980gravity,ne1980gravity,hehl1976general}.
Though $\Lambda$CDM model can account for most of the CMB observation very well, one of cosmic anomalies is the equationment of low-$l$ multipoles in the CMB angular power spectrum where $\Lambda$CDM's prediction do not fit the observation well. According to the recent Planck data, the normals of octopole plane, the quadrupole plane and the direction of dipole moment do not coincide evidently\cite{Perivolaropoulos2014}. One can not expect either the CMB spectrum obeys the cosmological principle in the CMB rest frame or the transformation from CMB rest frame to the peculiar motion frame is a simple Lorentz boost. There are many ways to compensate the low-$l$ multipole anomalies, e.g. assuming the anisotropic dark energy etc. All these solutions either sacrifice the cosmological principle or assuming non-Riemann geometry or nontrivial topology of the space-time of the universe\cite{Perivolaropoulos2014}.

\begin{figure}[!htbp]
	\centering	
	\subfigure[Illustration of the evolution of the universe and the primordial Lorentz violation]
	{
		\label{universe}
		\includegraphics[width=2in,height=1.3in]{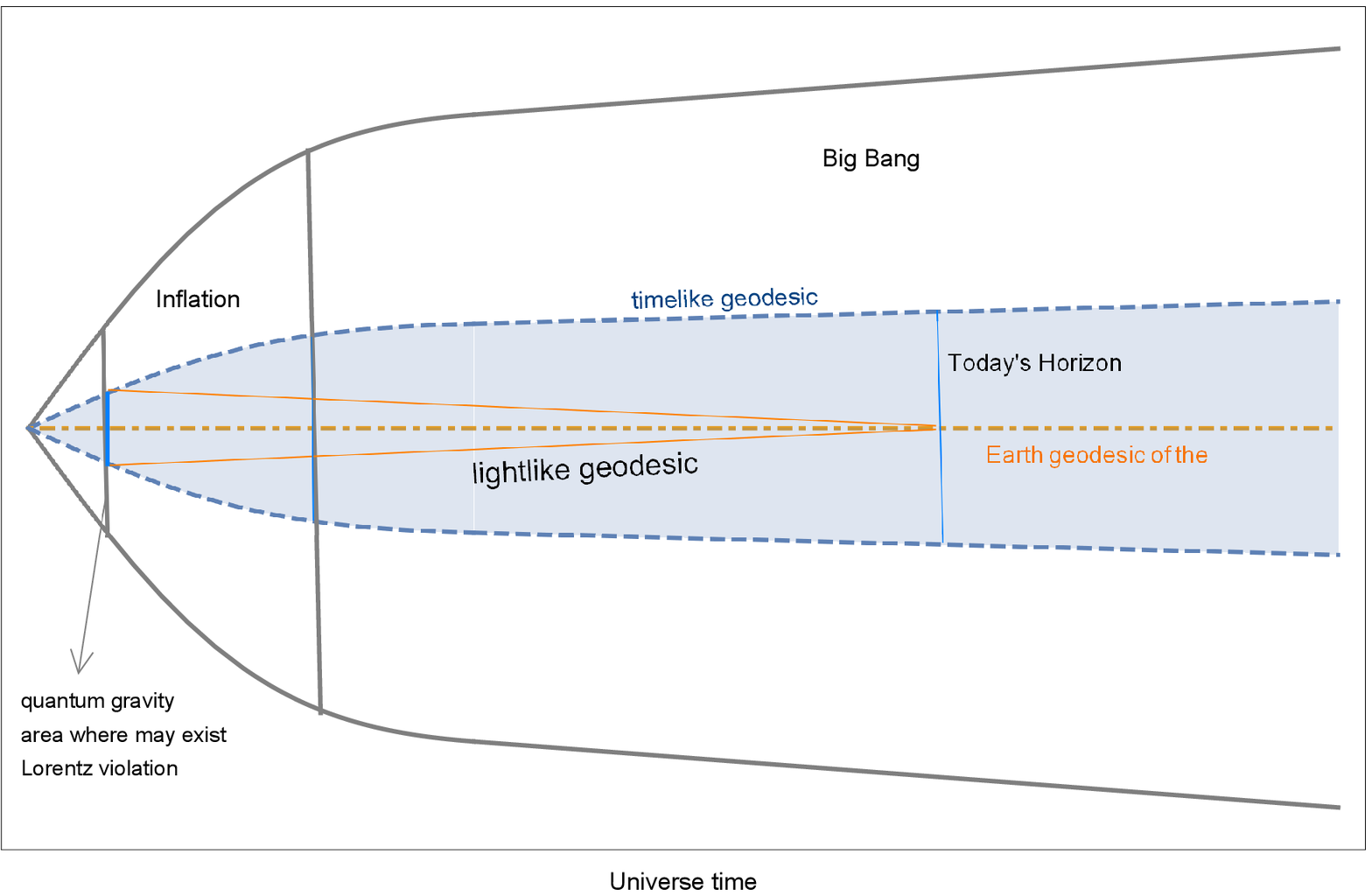}
	}
	\quad
	\subfigure[The Lorentz violation at the quantum gravity dominating era may be transformed into large scale by inflation]
	{
		\label{inflation}
		\includegraphics[width=2in,height=1.3in]{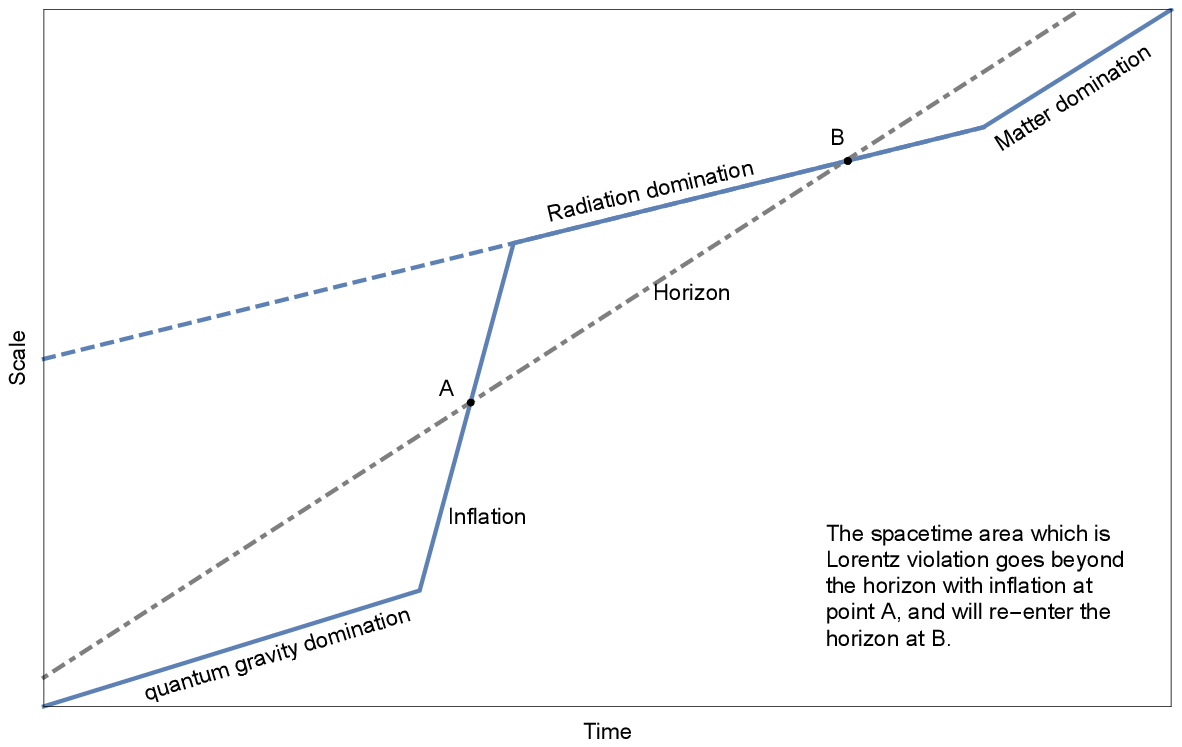}
	}
	\caption{The possible origin of large scale Lorentz violation}
\end{figure}

On the other hand, quantum gravity predicts the existence of minimum length scale $L_p$ with respect to the maximum energy scale $M_p$ and hence Lorentz invariance is lost at Planck scale, the main idea of doubly relativity\cite{Amelino-Camelia2002}. There are many approaches in quantum gravity leading to Lorentz violation\cite{kostelecky1989spontaneous, carroll2001noncommutative, alfaro2004alternative}. It is believed that our universe experiences a phase of inflation through which modern observable universe is expanded from a very tiny area of the primordial universe before inflation, when it is dominated by quantum gravity and therefore Lorentz violated, Fig.\ref{universe}. The primordial Lorentz violation(LV) may be transformed into large scale one by inflation. The scale of LV may exceed the horizon after inflation and re-enter the horizon along with the expansion of the universe, Fig.\ref{inflation}. Therefore we expect there may be LV at large scale esp. at the cosmic scale. To account for the large scale Lorentz violation(LSLV), one natural way is employing gauge principle in the construction of large scale effective gravity(LSEG).

The idea of large scale Lorentz violation is first proposed in 2015\cite{wu2015effective}. The framework of constructing large scale effective gravity with LV is build up with gauge principle via the equivalence principle by utilizing the constrain dynamics and very special relativity(VSR) symmetry as a modified gravity model\cite{wu2015effective,wu2016sim,yang2017VSR,wei2017E2}. The common feathers of these models are the non-trivial contortion distribution accompanied with the source matter energy-momentum distribution even with the scalar source matter, while contortion must be trivial and the space-time connection must be Levi-Civita one in GR in the case. The non-trivial contortion distribution contributes effectively a dark partner in addition to the matter source distribution. However, the VSR symmetry groups do not contain rotation symmetry totally, it is very hard to find a rotation symmetric solution in these models.
The dark partner may contribute dark energy effect in the cosmic scale or the dark matter effect at scale larger than the galaxy. The scenario of the origin of dark partner comes from large scale Lorentz violation is very different from traditional dark energy and dark matter scenario which suggests microscopic origin. However, the Lorentz invariance at scale smaller than the cosmic one may be broken partly and its contribution to the dark partner may be only partly. It is most probably that the LSLV effect is at the cosmic scale, the dark energy effect. We therefore concentrate on the connection between the LSLV and the dark energy effect.

\section{Gravity with Large Scale Lorentz Violation}
To construct a gravitation theory with LSLV, one can learn from the tetrad formalism of general relativity\cite{ramond2001field,Aldrovandi:2013wha}. 
The tetrad field ${h_a}^{\mu}$ is the coordinate transformation from anholonomic locally flat or free falling coordinates to a general holonomic coordinates. The commutator
\begin{equation}\label{commutator}
\left[h_a,h_b\right]={f^{c}}_{ab}h_{c}
\end{equation} 
for tetrad basis $h_{a}={h_a}^{\mu}\partial_{\mu}$ is non-trivial for anholonomic
coordinates in general. The metric tensor can be decomposed into the tetrad fields in the way $g_{\mu\nu}=\eta_{ab}{h^a}_{\mu}{h^b}_\nu$. The local Lorentz invariance of matter part action is guaranteed by introducing the Lorentzian gauge field ${{A}_{\mu }}=\dfrac{1}{2}{{A}^{ab}}_{\mu}{{S}_{ab}}$, which behaves as the connection and has a corresponding linear connection ${\Gamma^\rho}_{\nu\mu}={h_a}^\rho\left(\partial_{\mu }{h^a}_\nu+{A^a}_{b\mu}{h^b}_\nu\right)$, by the gauge principle with the substitution of ordinary derivative $\partial_{\mu}$ to the covariant derivative $\mathcal{D}_{\mu}=\partial_{\mu}-\dfrac{i}{2}{A^{ab}}_{\mu}S_{ab}$ in the matter part Lagrangian $\mathcal{L}_M$ in flat space-time background. Here ${S}_{ab}$ stand for the generators of Lorentz algebra.  The tetrad fields ${h_a}^{\mu}$ can also be regarded as gauge potential of the local translation of space-time in some sense. The Maurer-Cartan equation, $\left[ {{D}_{a}},{{D}_{b}}\right]={{T}_{ba}}^{p}{{D}_{p}}+\dfrac{i}{2}{{R}_{ba}}^{pq}{{S}_{pq}}$, supplies the curvature and torsion as the field strengths for Lorentz connection and tetrad field respectively.

The gravity part of the action consists of the Lorentz gauge field ${A^a}_{b\mu}$ and tetrad field ${h_a}^\mu$ can only be of Hilbert-Einstein type, 
\begin{equation}\label{HE}
S_{E}=\dfrac{1}{16\pi G}\int\mathrm d^{4}x\,hR
\end{equation}
where ${h=\det {h^a}_\mu}$, because Yang-Mills one gives neither the correct dimension of the gravitational coupling constant nor the Newtonian gravity limit. 
Denoting the energy momentum tensor and the angular momentum tensor of source matter distribution by ${\left(T_M\right)^a}_c$ and ${{\left(C_M\right)}_{ab}}^\mu$ respectively, the equations of motion (EoM) for ${A^a}_{b\mu}$,
\begin{equation}\label{EoMconnection}
\mathcal{D_{\nu}}\left(h{{h}_{a}}^{[\nu }{{h}_{b}}^{\mu]}\right)=8\pi G {{\left( {{C}_{M}} \right)}_{ab}}^{\mu }\,
\end{equation}
constrains the connections to be the Levi-Civita one in the case ${{\left( {{C}_{M}} \right)}_{ab}}^{\mu }=0$. The EoM for tetrad fields are the Einstein field equations 
\begin{equation}\label{EinsteinEQ}
{G^a}_b={{R}^{a}}_{b}-\frac{1}{2}R{{{\delta }^{a}}_{b}}=8\pi G{{{\left( {{T}_{M}} \right)}^{a}}_{b}} \ . 
\end{equation}

To modify GR in the LSLV at cosmic scale case, one can restrict the components of Lorentz gauge field ${A^a}_{b\mu}$ nontrivial only on $SO(3)$ generators to incorporate the Lorentz boost violation just as the VSR symmetry case in\cite{wu2015effective,wu2016sim,yang2017VSR,wei2017E2}. However, the boost transformation is not prohibited at the large scale actually for only the Lorentz boost transformation is violated. There are discussions on the modification of Lorentz algebra at quantum level by Hopf algebra or deformed Poincare algebra such as the  $\kappa$-Poincare etc. as well as other quantum gravity model like Horava-Lifshitz gravity in which the Lorentz boost is automatically violated. The restriction of Lorentz gauge field's components on boost generators vanishing is so strong that to induce the dynamics degenerating, especially for the equations of the Robertson-Walker like solution. We turn to seek a moderate way of introducing the Lorentz boost violation in the gauge gravity scheme by observing that the Lorentz gauge potentials transform as
\begin{equation}\label{transA}
{{A'}^{a}}_{b\mu }={{\Lambda }^{a}}_{c}\left( x \right){{A}^{c}}_{d\mu }{{\Lambda }_{b}}^{d}\left( x \right)+{{\Lambda }^{a}}_{c}\left( x \right){{\partial }_{\mu }}{{\Lambda }_{b}}^{c}\left( x \right)
\end{equation}
under local Lorentz transformation $\Lambda(x)$. It is obvious that ${A'^0}_{i\mu} ={{\Lambda }_{i}}^{j}\left( x \right){{A}^{0}}_{j\mu }$ for a rotation transformation ${\Lambda\in SO(3)}$. Hence the restriction to the Lorentz gauge potentials can proposed as
\begin{equation}\label{constraint}
\left({A^{0}}_{1\mu}\right)^2+\left({A^{0}}_{2\mu}\right)^2+\left({A^{0}}_{3\mu}\right)^2=\left(f_\mu (x)\right)^2
\end{equation}
where $\mu=t,r,\theta,\phi$ and $f_\mu(x)$ can be regarded as a spactime indexed vector field independent of tetrad indices , which measures the magnitude of the boost violation in some sense, and hence it is invariant under a local $SO(3)$ gauge transformation on tetrad fields ${h_a}^{\mu}$.

The action of LSLV effective gravity can then be given by the constrain dynamics with Lagrange multiplier items adding to the Hilbert-Einstein part ,
\begin{equation}\label{action}
\begin{split}
S_G=\frac{1}{16\pi G}\int\mathrm d^4x\,h \bigg(R
+\lambda^\mu \left(
\left({A^0}_{1\mu}\right)^2+\left({A^0}_{2\mu}\right)^2
+\left({A^0}_{3\mu}\right)^2-f_\mu^2\right)\bigg)
\end{split}
\end{equation}
where the repeated index $\mu$ of $\lambda^\mu$ and ${A^0}_{i\mu}$ means summation. 

The EoM of tetrad field keeps the form of Einstein equation,
\begin{equation}\label{G=T}
{G^a}_b={{R}^{a}}_{b}-\frac{1}{2}R{{{\delta }^{a}}_{b}}=8\pi G{{{\left( {{T}_{M}} \right)}^{a}}_{b}} \; ,
\end{equation}
where ${\left(T_M\right)^a}_b$ is the energy-momentum tensor. The connection here is no longer the Levi-Civita one because the boost violation constrain\eqref{constraint} changes the EoM for gauge potential in a given basis of tetrad ${h_{a}}^{\mu}$ to
\begin{equation}\label{connection}
{\mathcal{D}_{\nu }}\left( h{{h}_{0}}^{[\nu }{{h}_{i}}^{\mu ]} \right)\text{+}\frac{1}{2}\lambda^\mu h{{A}^{0}}_{i\mu }\text{=0} 
\end{equation}
where $i=1,2,3$ and the repeated $\mu$ does not mean summation just here for it is a result of variation respect to the square power of  ${A^0}_{i\mu}$ in eq.\eqref{action} and 
\begin{equation}\label{connection1}
{\mathcal{D}_{\nu }}\left( h{{h}_{i}}^{[\nu }{{h}_{j}}^{\mu ]} \right)\text{=}0	
\end{equation}
for the $i,j$ indices combination. Comparing eq.\eqref{connection} with eq. \eqref{EoMconnection}, the Lagrange multiplier term in eq. \eqref{connection} plays a similar role as the angular momentum tensor of source matter distribution ${{\left(C_M\right)}_{ab}}^\mu$ in eq.\eqref{EoMconnection} and hence induces the nontrivial torsion and contortion distribution in general. Eqs.\eqref{connection} and \eqref{connection1} will lead to Levi-Civita connection in GR in the absence of ${{\left(C_M\right)}_{ab}}^\mu$ for the scalar source matter distribution where the constrains don't exist and the Lagrange multipliers vanish. 

The spin connection can be decomposed into torsionless part and contorsion
\begin{equation}\label{decomposion}
{{A}^{a}}_{bc}={{\tilde{A}}^{a}}_{\; \;bc}+{{K}^{a}}_{bc}\, ,
\end{equation}
where
\begin{equation}\label{levicivita}
\widetilde{A}_{\; \;bc}^{a}=\dfrac{1}{2}\left( {{f}_{b}}{{^{a}}_{c}}+{{f}_{c}}{{^{a}}_{b}}-{{f}^{a}}_{bc} \right)
\end{equation} 
is the Levi-Civita connection, 
\begin{equation}\label{contortion}
{{K}^{a}}_{bc}=\dfrac{1}{2}( {{T}_{b}}{{^{a}}_{c}}+{{T}_{c}}{{^{a}}_{b}}\\-{{T}^{a}}_{bc} )
\end{equation}
is the contortion tensor and ${T^{a}}_{bc}$ is the torsion \cite{Aldrovandi:2013wha}. 
The tetrad EoM takes the form of \eqref{EinsteinEQ}. For only the
symmetric part of connection can affect motion of particle along
the geodesic equation, we decompose the curvature
with the help of decomposition of spin connection as
\begin{equation}
{{R}^{mn}}_{ab}={{\tilde{R}^{mn}}}_{\;\;\;\;\;\;ab}+{R_{K}}{{^{mn}}_{ab}}+{{R}_{CK}}{{^{mn}}_{ab}} \; ,
\end{equation}
where ${{\tilde{R}^{mn}}}_{\;\;\;\;\;\;ab}$ and ${{R}_{K}}{{^{mn}}_{ab}}$ are
the curvatures composed of torsion-free connection and contortion respectively,
while ${{R}_{CK}}{{^{mn}}_{ab}}$ contains cross terms of them. We can rewrite
\eqref{G=T} as
\begin{equation}\label{EoMsim2}
{{\tilde{R}}_{c}}^{\;\;a}-\frac{1}{2}{{\delta }_{c}}^{a}\tilde{R}=8\pi G{{\left( T_{eff}+{{T}_{M}} \right)}_{c}}^{a} 
\; ,
\end{equation}
where
\begin{equation}
{{\left( T_{eff} \right)}_{c}}^{a} = \dfrac{1}{8\pi G}\left( \dfrac{1}{2}{{\delta }_{c}}^{a}\left( {{R}_{K}}+{{R}_{CK}} \right)-\left( {{R}_{K}}{{_{c}}^{a}}+{{R}_{CK}}{{_{c}}^{a}} \right) \right) \; .
\label{T-Sim2}
\end{equation}

The force exerting on a particle moving in the gravitation field is supplied
by the Levi-Civita connection for it moves along the geodesic curve, the curvature of which satisfies
the Einstein field equation with the effective energy-momentum tensor
$T_{eff}$ of Eq.~(\ref{T-Sim2}) generated by an effective matter distribution
formally. It is worthy to note that $T_{eff}$ will
disappear if there is no matter distribution all over the space, {\it i.e.}
Minkowski spacetime is still the vacuum solution.

The effective energy-momentum tensor $T_{eff}$ contributes
to the gravitation in addition to matter contribution $T_{M}$
and appears as the dark partner of the matter distribution. Different source matter
distribution is expected to give rise different dark distribution.
At the cosmic scale, it is expected
to lead to the possible contribution to dark energy effectively.

With the decomposion ${{A^a}_{b\mu}}$ into Levi-Civita connection ${\widetilde A^a}_{\;\;b\mu}$ and contortion ${K^a}_{b\mu}$ in eq. \eqref{decomposion},
Eqs.\eqref{connection1} can be expressed in detail as
\begin{equation}\label{K}
\begin{split}
{K^0}_{12}={K^0}_{21},\;{K^1}_{23}=0,\;{K^2}_{12}=-{K^0}_{10},\;{K^3}_{13}=-{K^0}_{10},\\
{K^0}_{23}={K^0}_{32},\;{K^2}_{31}=0,\;{K^3}_{23}=-{K^0}_{20},\;{K^1}_{21}=-{K^0}_{20},\\
{K^0}_{31}={K^0}_{13},\;{K^3}_{12}=0,\;{K^1}_{31}=-{K^0}_{30},\;{K^2}_{32}=-{K^0}_{30}.
\end{split}
\end{equation}
and eqs.\eqref{connection} as
\begin{equation}\label{K0}
\begin{split}
&2{K^0}_{10}{h_0}^\mu+\left({K^0}_{22}+{K^0}_{33}\right){h_1}^\mu-\left({K^1}_{20}+{K^0}_{21}\right){h_2}^\mu +\left({K^3}_{10}-{K^0}_{31}\right){h_3}^\mu\\
&+\lambda^\mu\left({A^0}_{10}{h^0}_\mu +{A^0}_{11}{h^1}_\mu+{A^0}_{12}{h^2}_\mu+{A^0}_{13}{h^3}_\mu\right)=0 \\
&2{K^0}_{20}{h_0}^\mu+\left({K^1}_{20}-{K^0}_{12}\right){h_1}^\mu+\left({K^0}_{11}+{K^0}_{33}\right){h_2}^\mu-\left({K^2}_{30}+{K^0}_{32}\right){h_3}^\mu\\
&+\lambda^\mu\left({A^0}_{20}{h^0}_\mu +{A^0}_{21}{h^1}_\mu+{A^0}_{22}{h^2}_\mu+{A^0}_{23}{h^3}_\mu\right)=0\\
&2{K^0}_{30}{h_0}^\mu-\left({K^3}_{10}+{K^0}_{13}\right){h_1}^\mu+\left({K^2}_{30}-{K^0}_{23}\right){h_2}^\mu+\left({K^0}_{11}+{K^0}_{22}\right){h_3}^\mu\\
&+\lambda^\mu\left({A^0}_{30}{h^0}_\mu+{A^0}_{31}{h^1}_\mu+{A^0}_{32}{h^2}_\mu+{A^0}_{33}{h^3}_\mu\right)=0
\end{split}
\end{equation}
It can be expected in principle to solve all components of contortion ${K^a}_{b\mu}$ and tetrad field ${h_a}^{\mu}$ and the $\lambda$ multipliers out from eqs. \eqref{K} and \eqref{K0} as well as eq. \eqref{G=T} together with the constrains conditions\eqref{constraint} if the scalar matter energy-momentum distribution ${(T_M)^a}_b$ is given with ${{\left( {{C}_{M}} \right)}_{ab}}^{\mu }=0$ and the Lorentz violation parameters $f_\mu$ in eq. \eqref{constraint} are given. Since boost violation, the theory is not frame independent any more. One need to choose a frame to write down all the equations just like one does in a gauge theory where the EoM is gauge dependent and one needs to fixed a gauge. We choose the CMB rest frame as one in which the cosmology principle holds and suppose the theory stands in this frame. The cosmology principle guarantees FRW form of space-time metric 
\begin{equation}\label{FRW}
{\mathrm ds^2=\mathrm dt^2-a(t)^2\left(\dfrac{\mathrm dr^2}{1-kr^2}+r^2\mathrm d\theta^2+r^2\sin^2\theta\,\mathrm d\varphi^2\right)}\, ,
\end{equation}
from which it is easy to get the diagonal tetrad basis,
\begin{equation}\label{guage}
\begin{split}
h_0&=\frac{\partial}{\partial t},\;
h_1=\frac{\sqrt{1-kr^2}}{a(t)}\frac{\partial}{\partial r},\;
h_2=\frac{1}{a(t)r}\frac{\partial}{\partial\theta},\;
h_3=\frac{1}{a(t)r\sin\theta}\frac{\partial}{\partial\varphi};\;\\
h^0&=\mathrm dt,\;
h^1=\frac{a(t)}{\sqrt{1-kr^2}}\,\mathrm dr,\;
h^2=a(t)r\,\mathrm d\theta,\;
h^3=a(t)r\sin\theta\,\mathrm d\varphi\; ,
\end{split} 
\end{equation}
representing an isotropic observer in CMB Rest Frame. In this retrad, eqs. \eqref{K0} will be simplify as,
\begin{equation}\label{Kh0}
\begin{split}
2{K^0}_{10}{h_0}^t + {\lambda ^t}{K^0}_{10}{h^0}_t = 0\\
2{K^0}_{20}{h_0}^t+{\lambda ^t}{K^0}_{20}{h^0}_t=0\\
2{K^0}_{30}{h_0}^t+{\lambda ^t}{K^0}_{30}{h^0}_t=0
\end{split}
\end{equation}
\begin{equation}\label{Kh1}
\begin{split}
&\left( {{K^0}_{22} + {K^0}_{33}} \right){h_1}^r + {\lambda ^r}{A^0}_{11}{h^1}_r = 0\\
&\left( {{K^1}_{20} - {K^0}_{12}} \right){h_1}^r + {\lambda ^r}{K^0}_{21}{h^1}_r = 0\\
&\left( {{K^3}_{10} + {K^0}_{13}} \right){h_1}^r - {\lambda ^r}{K^0}_{31}{h^1}_r = 0
\end{split}
\end{equation}
\begin{equation}\label{Kh2}
\begin{split}
&\left( {{K^1}_{20} + {K^0}_{21}} \right){h_2}^\theta  - {\lambda ^\theta }{K^0}_{12}{h^2}_\theta  = 0\\
&\left( {{K^0}_{11} + {K^0}_{33}} \right){h_2}^\theta  + {\lambda ^\theta }{A^0}_{22}{h^2}_\theta  = 0\\
&\left( {{K^2}_{30} - {K^0}_{23}} \right){h_2}^\theta  + {\lambda ^\theta }{K^0}_{32}{h^2}_\theta  = 0\\
\end{split}
\end{equation}
and,
\begin{equation}\label{Kh3}
\begin{split}
&\left( {{K^3}_{10} - {K^0}_{31}} \right){h_3}^\phi  + {\lambda ^\phi }{K^0}_{13}{h^3}_\phi  = 0\\
&\left( {{K^2}_{30} + {K^0}_{32}} \right){h_3}^\phi  - {\lambda ^\phi }{K^0}_{23}{h^3}_\phi  = 0\\
&\left( {{K^0}_{11} + {K^0}_{22}} \right){h_3}^\phi  + {\lambda ^\phi }{A^0}_{33}{h^3}_\phi  = 0
\end{split}
\end{equation}
We need to combine eqs.\eqref{Kh0}---\eqref{Kh3} together with eq.\eqref{G=T} to find the solution for ${K^\mu}_{ab}$. Noting that, the ideal fluid $T_M$ in eq.\eqref{G=T} requires ${{G^a}_{b}=0,\;\forall a\not=b}$, which will lead to,
\begin{equation}\label{K0i0}
{K^0}_{i0}=0,\; i=1,2,3
\end{equation}
It makes the value of $\lambda^t$ not able to be fixed by eq.\eqref{Kh0}. The other values of $\lambda^\mu$ can be solved out by eq.\eqref{Kh1}---\eqref{Kh3},
\begin{equation}\label{lambda}
\begin{split}
&{\lambda ^r} =- \frac{{\left( {{K^0}_{22} + {K^0}_{33}} \right)}}{{{A^0}_{11}}}\\
&{\lambda ^\theta } =-\frac{{\left( {{K^0}_{11} + {K^0}_{33}} \right)}}{{{A^0}_{22}}}\\
& {\lambda ^\phi } =-\frac{{\left( {{K^0}_{11} + {K^0}_{22}} \right)}}{{{A^0}_{33}}}
\end{split}
\end{equation} 
However, the values of ${K^0}_{i0}=0,\; i=1,2,3$ should be accidental duo to the RW metric scenario. Any kind of cosmic perturbation away from RW metric can destroy the highly symmetrical solution of \eqref{K0i0}. Let's consider a special kind of perturbation to the RW metric which keeps the tetrad in diagnal form, i.e. ${h'^a}_{\mu} \propto {\delta^a}_{\mu}$ . The benefit is that eqs. \eqref{Kh0} will have a similar form for the perturbed ${K'^0}_{i0}={K^0}_{i0}+\delta {K^0}_{i0}$ and ${h'^a}_{\mu} ={h^a}_{\mu} +\delta {h^a}_{\mu}$, i.e.
\begin{equation}\label{Kh0prime}
\begin{split}
2{K'^0}_{10}{h'_0}^t + {\lambda ^t}{K'^0}_{10}{h'^0}_t = 0\\
2{K'^0}_{20}{h'_0}^t+{\lambda ^t}{K'^0}_{20}{h'^0}_t=0\\
2{K'^0}_{30}{h'_0}^t+{\lambda ^t}{K'^0}_{30}{h'^0}_t=0
\end{split}
\end{equation}
The requiement of ${{G^a}_{b}=0,\;\forall a\not=b}$ will give the $\delta {K^0}_{10}$ dependence on $\delta {h^a}_{\mu}$ and leads to 
\begin{equation}
{\lambda ^t} =- \frac{{2h'_0}^t}{{h'^0}_t}\\
\end{equation}
Consider the metric perturbation of the form
\begin{equation}\label{metricpert}
{d}{{s}^{2}}={{\left( 1+\varepsilon \left( r \right) \right)}^{2}}{d}{{t}^{2}}-a{{(t)}^{2}}{{\left( 1-\delta \left( r \right) \right)}^{2}}\left( \frac{\text{d}{{r}^{2}}}{1-k{{r}^{2}}}+{{r}^{2}}\text{d}{{\Omega }^{2}} \right)\\,
\end{equation}
where $\varepsilon \left( r \right)$ and $\delta \left( r \right)$ are perturbations to potential and curvature respectively. It can be shown that ${K^0}_{10}\sim \varepsilon' \left( r \right)$ and hence
\begin{equation}\label{lbdat}
{\lambda ^t} =- \frac{2}{(1+\varepsilon (r) )^{2}}\xrightarrow{\varepsilon \to 0}-2\\
\end{equation}
The cosmological principle also requires all cosmic physical quantities depend only on cosmic time $t$, and hence the solution
\begin{equation}\label{K0ii}
{K^0}_{11}={K^0}_{22}={K^0}_{33}=\mathscr K(t)
\end{equation}
guarantees the requirement of ${G^1}_1={G^2}_2={G^3}_3$ from eq.\eqref{G=T} by the ideal fluid assumption of the cosmic matter source.
Finally, all the components of contortion can be determined to be trivial except ${{K^0}_{11},{K^0}_{22},{K^0}_{33}}$(and their symmetric partners ${{K^1}_{01},{K^2}_{02},{K^3}_{03}}$ obviously) which can be expressed in terms of $f_\mu(x)$ nd vice versa by
\begin{equation}\label{fmu}
\begin{split}
\left(f_t,f_r,f_\theta,f_\varphi\right)=\left(a(t)\mathscr K(t)+\dot a(t)\right)\cdot\left(0, \frac{1}{\sqrt{1-kr^2}}, r, r\sin\theta\right)
\end{split} 
\end{equation}
It should be noted that $f_\mu(x)$ introduced in eq.\eqref{constraint} are the Lorentz violation parameters which are not able to be predicted from the present gravitation model with Lorentz violation. A more fundamental theory of quantum gravity would have the ability to predict the Lorentz violation measure $f_\mu(x)$ via the mechanism of large scale Lorentz violation proposed in this paper. For the lack of appropriate quantum gravity model to predict $f_\mu(x)$ or equivalently $\mathscr K(t)$ from eq. \eqref{fmu}, we have to turn to seek some phenomenological approximations about $\mathscr K(t)$. 

The special form of $f_\mu(x)$ in eq.\eqref{fmu} seems that there is only one  degree of freedom to choose $f_\mu(x)$. It need to be noted that all the explicit form of equations concerning connection and hence contortion as well as the constrain eq.\eqref{constraint} are all frame dependent. The degrees of freedom of $f_\mu(x)$ actually hide in the local Lorentz boost transformation and the hence there are four independent choice of $f_\mu(x)$ for different frames choice. Our choice of frame just make the $f_\mu(x)$ to take the simplest form duo to symmetrical reason.

The Lagrange multiplier field $\lambda^\mu$ are fixed as,
\begin{equation}
\lambda^t=-2,\;\; \lambda^r=\lambda^\theta=\lambda^\phi=-\frac{2\mathscr K}{\dfrac{\dot a}{a}+\mathscr K}.
\end{equation}


\section{Cosmology of Large Scale Lorentz Violation Gravity}
Denote the covariant derivative and the Einstein tensor generated by Levi-Civita connection $\widetilde A_\mu$ as $\widetilde\nabla$ and $\widetilde{G}^a_{\,\;c}$ respectively, 
${G^a}_c$ can be decomposed into,
\begin{equation}
\begin{split}
{G^a}_c&=\widetilde{G}^a_{\,\;c}+2\left(\widetilde\nabla_{[c}{K^{ab}}_{b]}+{K^a}_{e[c}{K^{eb}}_{b]}-\frac{1}{2}\left(\widetilde\nabla_{d}{K^{db}}_{b}+{K^d}_{e[d}{K^{eb}}_{b]}\right){\delta^a}_c\right)
\end{split}
\end{equation}
by the decomposition \eqref{decomposion}. We can rewrite the tetrad field equation of the form \eqref{G=T} into 
\begin{equation}\label{dark energy}
\begin{split}
{\widetilde{R}^{a}}_{\,\;c}-\frac{1}{2}\widetilde{R}{\delta^a}_c=8\pi G{\left(T_M+T_\Lambda\right)^a}_c
\end{split}
\end{equation}
where ${{{T_\Lambda}^a}_c=\dfrac{1}{8\pi G}\left(\widetilde{G}^a_{\,\;c}-{G^a}_c\right)}$ can be regarded as the dark partner of the source matter distribution generated by the large scale Lorentz violation effect.
If $T_\Lambda$ can lead to the accelerating expansion of the universe, we can make the conclusion that the large scale Lorentz violation is the origin of the dark energy. 

With the results from eqs.\eqref{K},\eqref{K0i0} and \eqref{K0ii}, the effective dark partner for ideal fluid ${{[T_M]^a}_c=Diag(\rho,-p,-p,-p)}$ can be derived from \eqref{G=T}  as
\begin{equation}\label{Tlam}
{\left[{T_\Lambda}\right]^a}_c=Diag(\rho_\Lambda,-p_\Lambda,-p_\Lambda,-p_\Lambda)
\end{equation}
where ${\rho_\Lambda=-\dfrac{1}{8\pi G}\left(3\mathscr K^2+6\mathscr K\dfrac{\dot a}{a}\right)}$ and ${p_\Lambda=\dfrac{1}{8\pi G}\left(\mathscr K^2+4\mathscr K\dfrac{\dot a}{a}+2\dot{\mathscr K}\right)}$ respectively.
So is the modified Friedmann Equation,
\begin{equation}\label{fri1}
\begin{split}
&\left(\frac{\dot a}{a}\right)^2+\frac{k}{a^2}+2\mathscr K\frac{\dot a}{a}+\mathscr K^2=\frac{8\pi G}{3}\rho\\
&\ddot a=-4\pi G\cdot a\left(p+\frac{\rho}{3}\right)-\frac{\mathrm d}{\mathrm dt}\left(a\mathscr K\right)\; ,
\end{split}
\end{equation}
while one in $\Lambda$CDM model as a comparison,
\begin{equation}\label{fri2}
\begin{split}
&\left(\frac{\dot a}{a}\right)^2+\frac{k}{a^2}-\frac{1}{3}\Lambda=\frac{8\pi G}{3}\rho\\
&\ddot a=-4\pi G\cdot a\left(p+\frac{\rho}{3}\right)+\frac{1}{3}a\Lambda\; .
\end{split}
\end{equation}
By the modified Friedmann Equation\eqref{fri1}, the accelerating expansion solution of the universe corresponds the condition  ${4\pi G\cdot a\left(p+\dfrac{\rho}{3}\right)+\dfrac{\mathrm d}{\mathrm dt}\left(a\mathscr K\right)<0}$. It should be noted that $\mathscr K$ in eq.\eqref{fri1} can not be determined by eq.\eqref{fri1} even if the equation of state(EoS) of the cosmic media is given.  From the discussion concerning eq.\eqref{fmu}, $\mathscr K$ can be expressed in terms of $f_\mu$ which should be given by more fundamental theory, quantum gravity, or as an observational input just like $\Lambda$ in eq.\eqref{fri2} of $\Lambda$CDM model. Since cosmology based on $\Lambda$CDM model is already very successful in many aspects, the LSLV modification should take $\Lambda$CDM model as a good approximation. However, the comparison of eq.\eqref{fri1} and eq.\eqref{fri2} reveals the modified Friedmann Equation can not be equivalent to the one in $\Lambda$CDM model. We will hence make propositions in three cases to fix the evolution of $\mathscr K$ as close to $\Lambda$CDM model as possible.

\textbf{Case A: } By comparing eq.\eqref{fri1} and eq.\eqref{fri2}, we can make the proposition
\begin{equation}\label{case1}
-\frac{\mathrm d}{\mathrm dt}\left(a\mathscr K\right)=\frac{1}{3}a\Lambda
\end{equation}
and get the initial condition,
\begin{equation}\label{origin1}
2\mathscr K(t_0)\frac{\dot a(t_0)}{a(t_0)}+\mathscr K(t_0)^2=-\frac{1}{3}\Lambda
\end{equation}
where $t_0\simeq H_0^{-1}$ is the moment at present , or the age of universe now and  $H_0$ is Hubble constant. From eq.\eqref{origin1} one can get ${\mathscr K(t_0)=H_0\left(\pm\sqrt{1-\Omega_{\Lambda}}-1\right)}\simeq -0.465H_0$ or $-1.535H_0$ respectively, where ${\Omega_\Lambda=\dfrac{\Lambda}{3H_0^2}}$.
The expressions of conservation law for cosmic media in $\Lambda$CDM and LSLV model can be derived from eq.\eqref{fri1} and eq.\eqref{fri2},
\begin{itemize}
	\item $\Lambda$CDM:
	\begin{equation}
	\dot\rho+3\left(\rho+p\right)\frac{\dot a}{a}=0,
	\end{equation}
	\item LSLV:
	\begin{equation}\label{matter density}
	\quad\dot\rho+3\left(\rho+p\right)\frac{\dot a}{a}+\left(3p+\rho\right)\mathscr K=0\; .
	\end{equation}
\end{itemize}
The general solution of eq.\eqref{matter density} can be given by,
\begin{equation}\label{LV rho}
\begin{split}
\frac{8\pi G}{3H_0^2}\rho=\Omega_m\left(\frac{a_0}{a}\right)^3\exp{\left(\int^{t_0}_t\mathscr K\mathrm d\tau\right)}+\Omega_r\left(\frac{a_0}{a}\right)^4\exp{\left(2\int^{t_0}_t\mathscr K\mathrm d\tau\right)}
\end{split}
\end{equation}
where $a_0=a(t_0)$ and $\Omega_m$ is density parameter for dust($p=0$) while $\Omega_r$  is one for radiation($p=\rho/3$) respectively. 

\textbf{Case B: }Suppose the EoS of cosmic media is ${p(t)=w(t)\rho(t)}$, \eqref{fri1} and \eqref{fri2} can be converted to,
\begin{itemize}
	\item $\Lambda$CDM:
	\begin{equation}\label{ddot a CDM}
	\frac{\ddot a}{a}+\frac{3w+1}{2}\frac{\dot a^2+k}{a^2}=\frac{w+1}{2}\Lambda
	\end{equation}
	\item LSLV:
	\begin{equation}\label{ddot a LV}
	\begin{split}
	\frac{\ddot a}{a}+\frac{3w+1}{2}\frac{\dot a^2+k}{a^2}=-\dot{\mathscr K}-\frac{3w+1}{2}\mathscr K^2-(3w+2)\frac{\dot a}{a}\mathscr K	
	\end{split}
	\end{equation}
\end{itemize}

We can make a proposition by requiring 
\begin{equation}\label{case2}
\dot{\mathscr K}+\frac{3w+1}{2}\mathscr K^2+(3w+2)\frac{\dot a}{a}\mathscr K=-\frac{w+1}{2}\Lambda\; ,
\end{equation}
which relates contortion to $w(t)$. The initial conditions can be taken as the same as {\bf Case A}, ${\mathscr K(t_0)=-0.465H_0}$ or $-1.535H_0$ respectively.

\textbf{Case C: }  To mimic the contribution by cosmological constant to cosmology with one by contortion, we make an alternative proposition by requiring ${{T_\Lambda}^a}_b\propto {\delta^a}_b$, or ${{{{T_\Lambda}^0}_0={{T_\Lambda}^1}_1}={{T_\Lambda}^2}_2={{T_\Lambda}^3}_3}$, which leads to the equation satisfied by contortion,
\begin{equation}\label{case3}
\dot{\mathscr K}=\mathscr K^2+\mathscr K\frac{\dot a}{a}\;.
\end{equation}
The initial condition is obvious, ${{{T_\Lambda}^a}_b(t_0)=\dfrac{1}{8\pi G}\Lambda{\delta^a}_b}$, which is the same as one given by \eqref{origin1} exactly, i.e., $\mathscr K(t_0)=-0.465H_0$ or $-1.535H_0$.

We will concentrate on ${k=0}$ case of the metric here for our universe is spatially flat by observation. The extension to ${k= \pm 1}$ is straightforward and will be presented elsewhere. The evolution of $H(t)$ and $\mathscr K(t)$ can be determined by the equations for Hubble parameters $H(t)$ derived from \eqref{fri1} together with eq.\eqref{case1}, eq.\eqref{case2} and eq.\eqref{case3} and with the initial condition $H(t_0)=H_0$. 
Table \ref{model} summarizes the cases discussed above.
\begin{table}[!htbp]
	\centering
	\caption{Proposed Models of LSLV Cosmology}\label{model}
	\begin{tabular}{ccc}
		\toprule
		&Propositions on $\mathscr K(t)$
		&Values of $\mathscr K(t_0)$\\
		\midrule
		{\bf Case A-1}
		&\multirow{2}{*}{${\dot{\mathscr K}+H\mathscr K=-\dfrac{1}{3}\Lambda}$}
		&$-0.465H_0$\\
		{\bf Case A-2}
		&
		&$-1.535H_0$\\
		\hline
		{\bf Case B-1}
		&\quad\multirow{2}{*}{$\quad\dot{\mathscr K}+\dfrac{3w+1}{2}\mathscr K^2+(3w+2)H\mathscr K=-\dfrac{w+1}{2}\Lambda$}
		&$-0.465H_0$\\
		{\bf Case B-2}
		&
		&$-1.535H_0$\\
		\hline
		{\bf Case C-1}
		&\multirow{2}{*}{${\dot{\mathscr K}=\mathscr K^2+\mathscr KH}$}
		&$-0.465H_0$\\
		{\bf Case C-2}
		&
		&$-1.535H_0$\\
		\bottomrule
	\end{tabular}
\end{table}

\begin{figure}[!htbp]
	\centering
	\subfigure[$H$'s evolution from ${a\simeq 0}$ to ${a=2a_0}$.]
	{
		\label{LV1}
		\includegraphics[width=2in,height=1.3in]{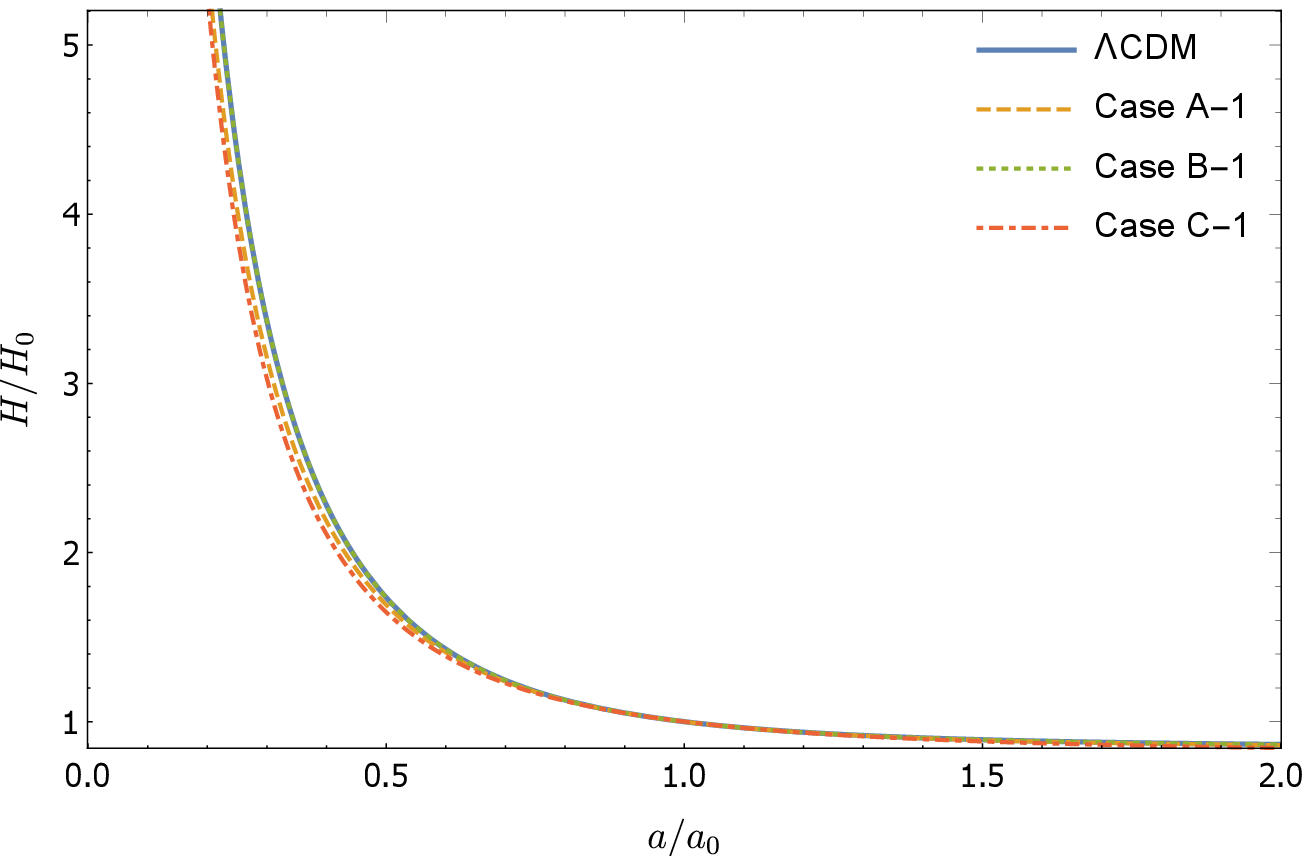}
	}
	\quad
	\subfigure[$H$'s evolution from ${a=a_0}$ to ${a=20a_0}$.]
	{
		\label{LV2}
		\includegraphics[width=2in,height=1.3in]{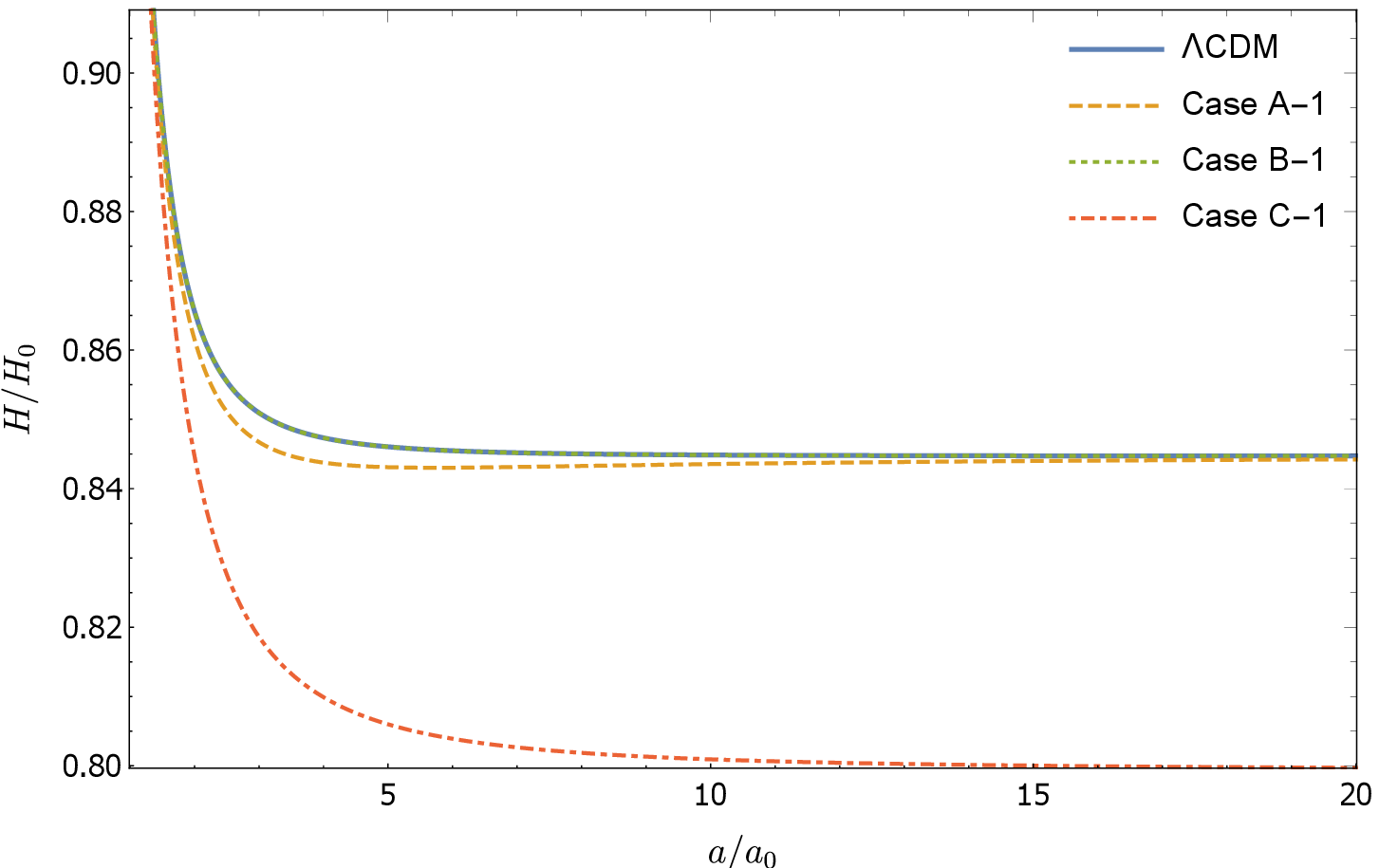}
	}
	\label{bigK}
	\caption{Evolutions of $H$ and $\mathscr K$ between LSLV model and $\Lambda$CDM model, with initial condition ${\mathscr K(t_0)=-0.465H_0}$}\label{fig1}
\end{figure}

\begin{figure}[!htbp]
	\centering
	\subfigure[$H$'s evolution from ${a\simeq 0}$ to ${a=2a_0}$.]
	{
		\label{LV3}
		\includegraphics[width=2in,height=1.3in]{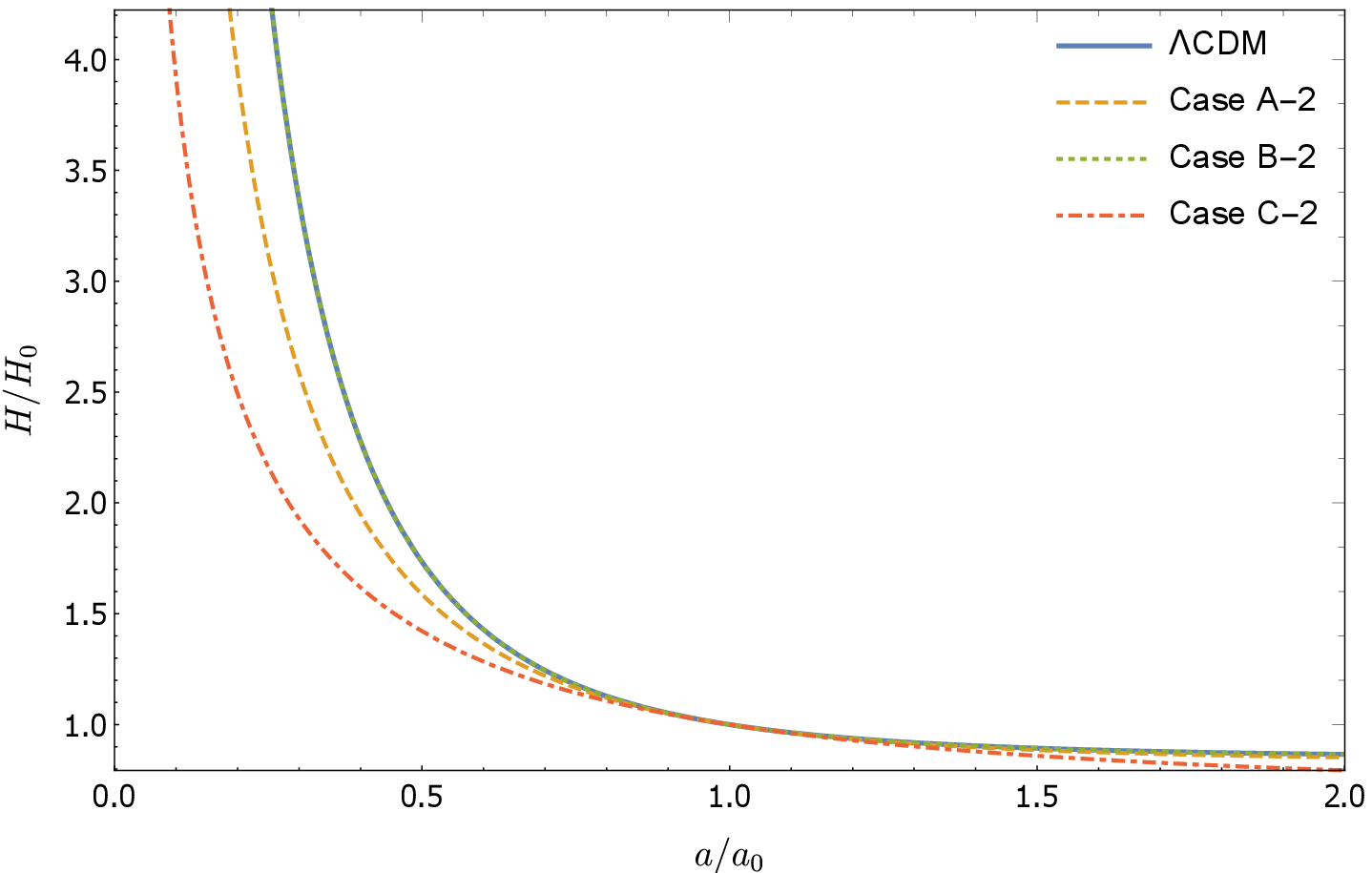}
	}
	\quad
	\subfigure[$H$'s evolution from ${a=a_0}$ to ${a=20a_0}$.]
	{
		\label{LV4}
		\includegraphics[width=2in,height=1.3in]{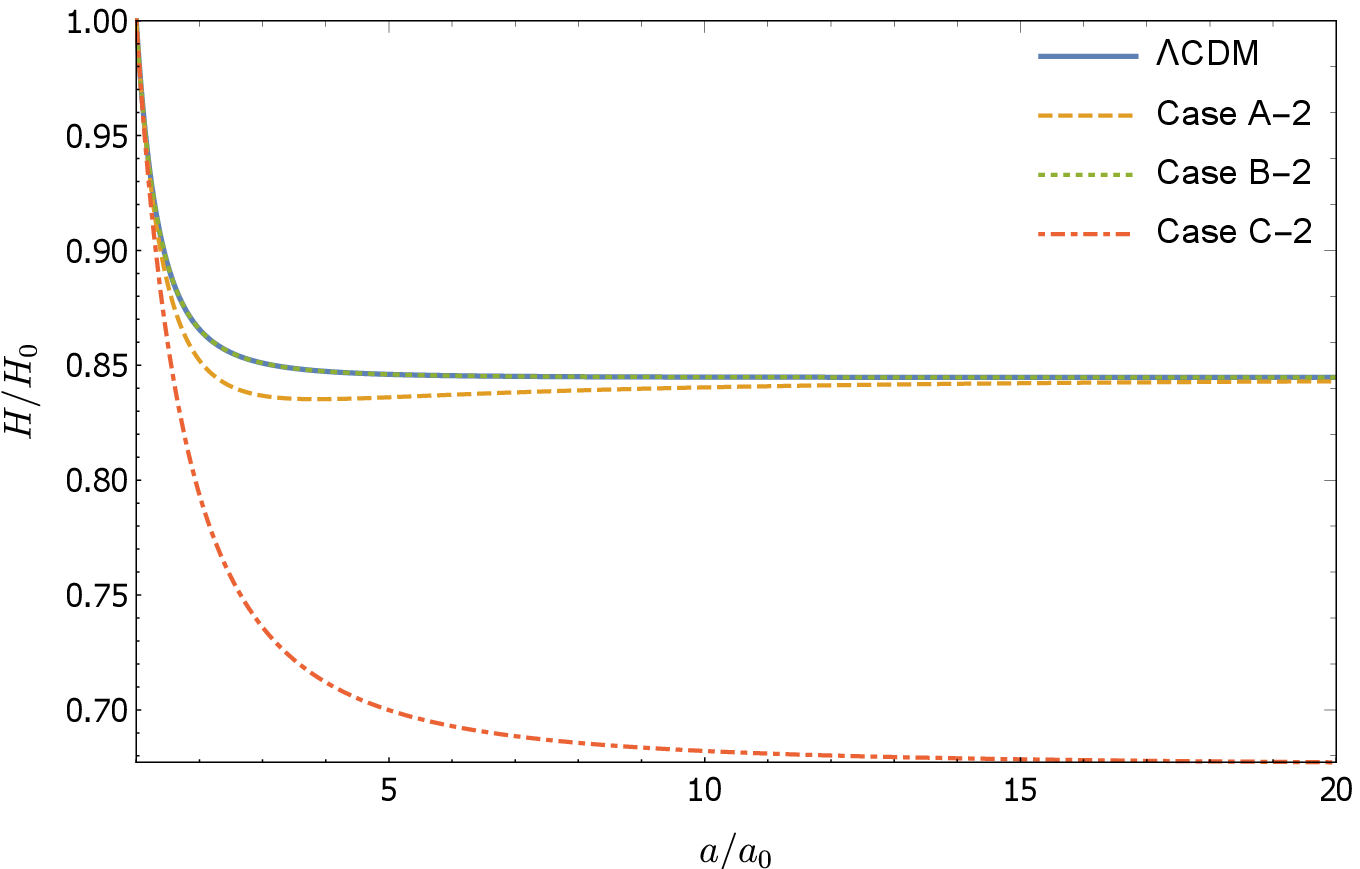}
	}
	\caption{Evolutions of $H$ and $\mathscr K$ between LSLV model and $\Lambda$CDM model with initial conditions ${\mathscr K(t_0)=-1.535H_0}$}\label{fig2}
\end{figure}

The two initial values of $\mathscr K(t_0)$, $-0.465H_0$ and $-1.535H_0$ correspond to two different solutions of the modified Friedmann Equation\eqref{fri1}, ${H=\sqrt{\dfrac{8\pi G}{3}\rho}-\mathscr K}$ and ${H=-\sqrt{\dfrac{8\pi G}{3}\rho}-\mathscr K}$ respectively. Obviously, the former is closer to result of $\Lambda$CDM model ${H=\sqrt{\dfrac{8\pi G\rho+\Lambda}{3}}}$. 
The evolution trend of $H(a)$ and $\mathscr K(a)$ versus scale factor $a$ in these different models are presented in Fig.\ref{fig1} and Fig.\ref{fig2}. 
We can set ${w(t)\simeq 0}$ for the late universe which is dominated by cold matter with $\Omega_m\gg\Omega_r$ in eq.\eqref{matter density}.

The evolution curves of $H$ in {\bf Case B} and in $\Lambda$CDM are coincided basically, because {\bf Case B} takes both matter and dark partner from LSLV into account. 
As can be observed, the evolution of all of these models is very close in the period near Hubble time, and {\bf Case A-1} approaches exactly the same with $\Lambda$CDM model in far future. If $\mathscr K(t_0)$ takes value of $-1.535H_0$, the differences of $H(a)$ among these models become bigger than Fig. \ref{LV1} near Hubble time. However trends of Fig.\ref{LV4} and Fig.\ref{LV2} are the same in the far future.

\begin{figure}[!htbp]
	\centering
	\subfigure[Evolutions of $\mathscr K$ with initial conditions ${\mathscr K(t_0)=-0.465H_0}$]
	{
		\label{LVK1}
		\includegraphics[width=2in,height=1.3in]{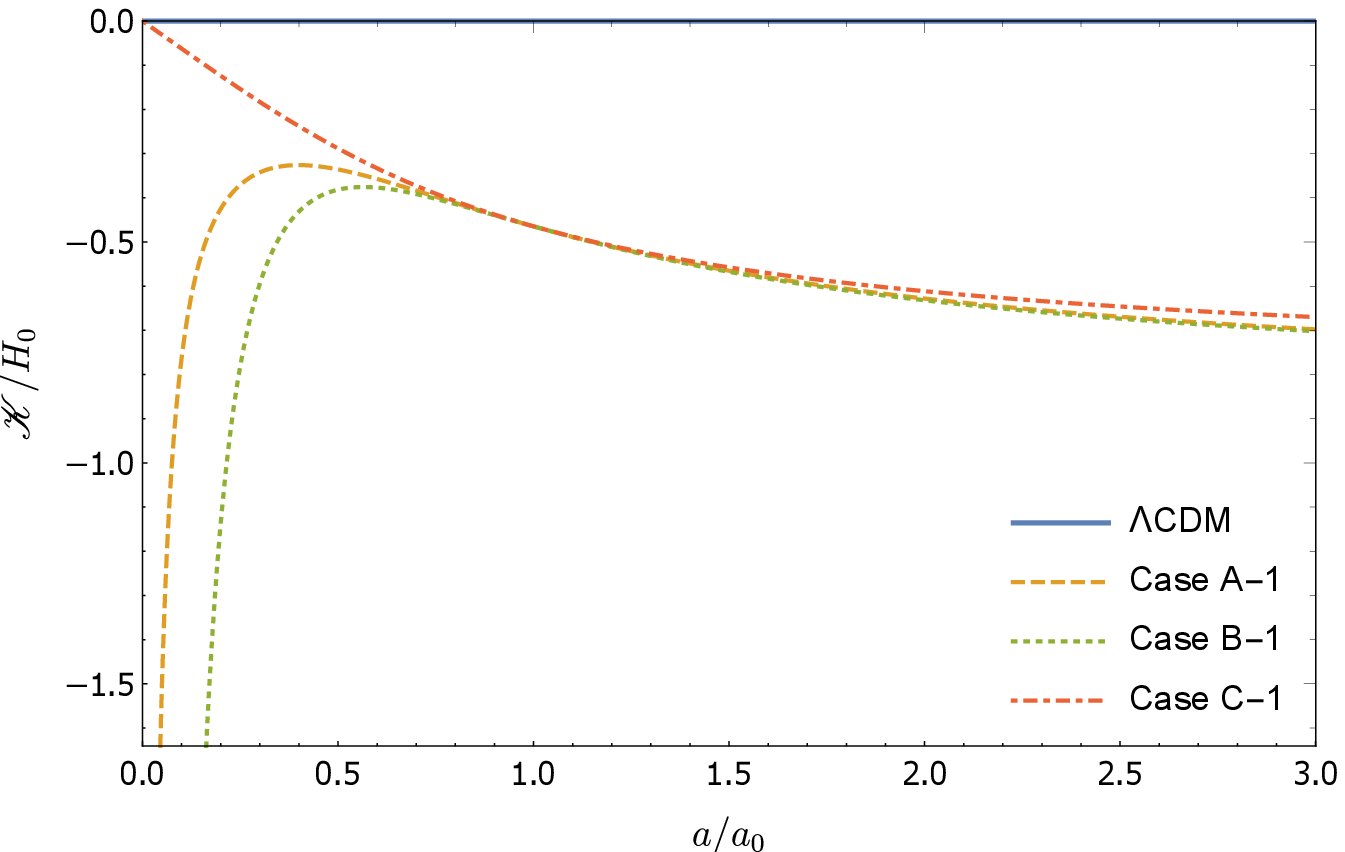}
	}
	\quad
	\subfigure[Evolutions of $\mathscr K$ with initial conditions ${\mathscr K(t_0)=-1.535H_0}$]
	{
		\label{LVK2}
		\includegraphics[width=2in,height=1.3in]{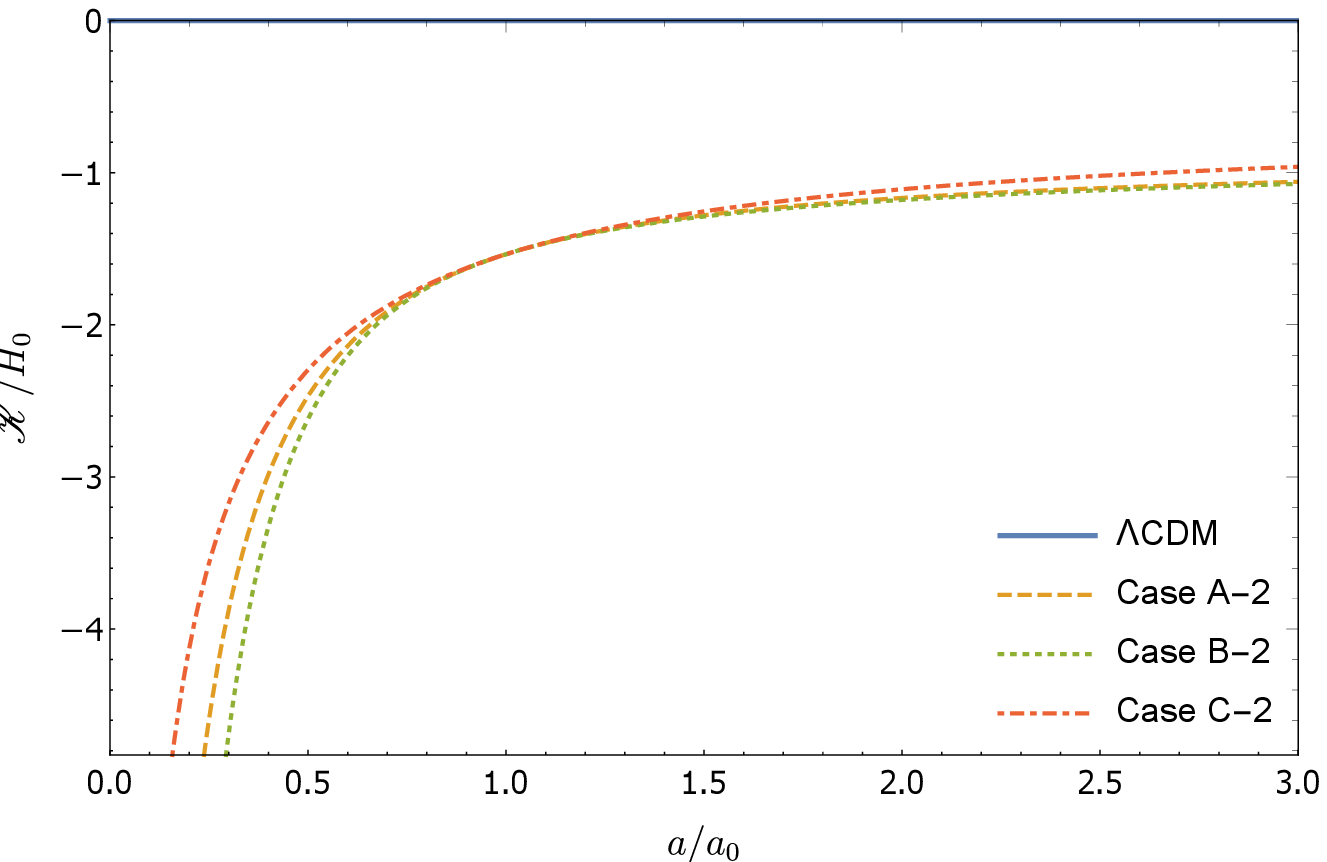}
	}
	\caption{$\mathscr K$'s evolution from $a\simeq 0$ to $a=3a_0$.}
\end{figure}

In the case of nontrivial contortion, the space-time is a Riemann-Cartan manifold on which the curve $x(\xi)$ satisfying autoparallel condition $\dfrac{\mathrm d^2x^\rho}{\mathrm d\xi^2}+{\Gamma^\rho}_{(\mu\nu)}\dfrac{\mathrm dx^\mu}{\mathrm d\xi}\dfrac{\mathrm dx^\nu}{\mathrm d\xi}=0$ is different from the geodesic one satisfying $\dfrac{\mathrm d^2x^\rho}{\mathrm d\xi^2}+\widetilde{\Gamma}^\rho_{\;\;\mu\nu}\dfrac{\mathrm dx^\mu}{\mathrm d\xi}\\\dfrac{\mathrm dx^\nu}{\mathrm d\xi}=0$. The world line for a free falling particle can be determined by the Hamilton principle which gives exactly the geodesic curve. The world line of a photon along which the proper time vanishes identically is just the light-like geodesic curve rather than the autoparallel one as in some literature\cite{zhang2017poincar}.
A particle feels the gravity via the curvature of space-time generated by Levi-Civita connection. Contortion contributes to the gravity by the effective dark partner energy-momentum tensor as the source of gravity instead of by the Riemann-Cartan geometry directly as is shown in eq.\eqref{dark energy}.
Therefore the redshift formula is the same as in $\Lambda$CDM model,
\begin{equation}
1+z=\frac{a_0}{a}
\end{equation}

\begin{figure}[!htbp]
	\centering
	\subfigure[Comparison of distance magnitudes. Curves of different models are hard to distinguish.]
	{	
		\label{m1}
		\includegraphics[width=2in,height=1.3in]{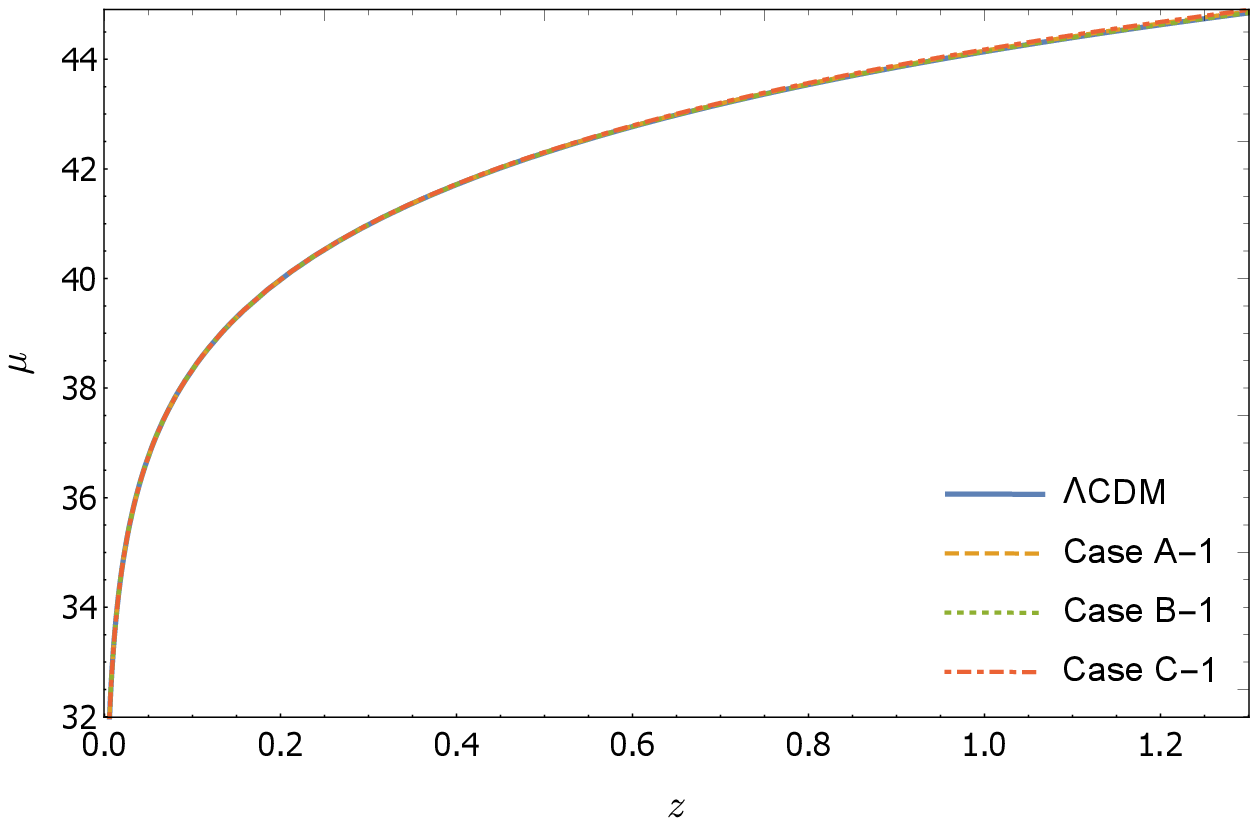}	
	}
	\quad
	\subfigure[Comparison of luminosity distance of these models.]
	{	
		\label{dL1}
		\includegraphics[width=2in,height=1.3in]{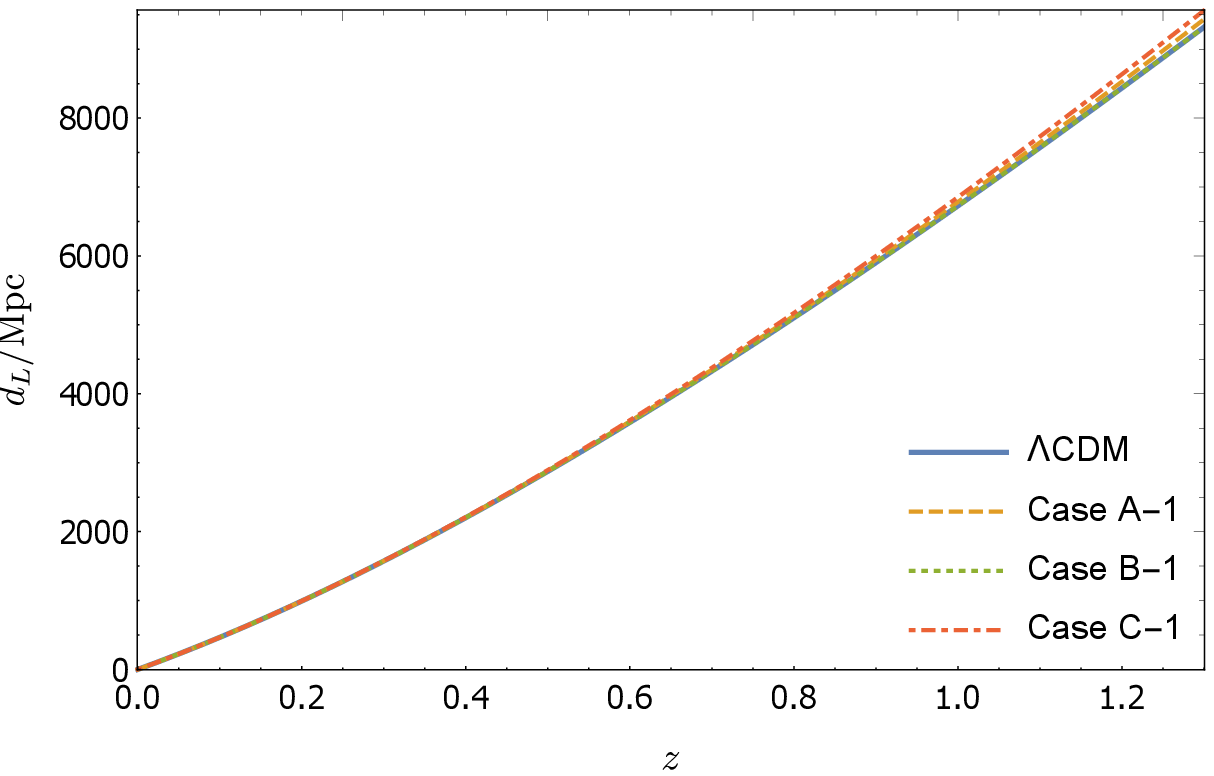}		
	}
	\caption{${\mathscr K(t_0)=-0.465H_0}$ Case}\label{observe1}
\end{figure}

By definition of luminosity distance $d_L$ (see \cite{Weinberg:2008zzc}) we can get,
\begin{equation}\label{hzdl}
d_L(z)=\left(1+z\right)\int^z_0\frac{1}{H(z')}\,\mathrm dz'\quad\left(\mathrm{for}\; k=0\right)
\end{equation}
and 
\begin{equation}\label{dtdz}
\frac{\mathrm dt}{\mathrm dz}=-\frac{1}{1+z}\,\frac{\mathrm d}{\mathrm dz}\left(\frac{d_L}{1+z}\right)
\end{equation}
With eq.\eqref{hzdl} and eq.\eqref{dtdz}, we can convert eq.\eqref{fri1} and propositions in table.\ref{model} to equations for $d_L(z)$ and $\mathscr K(z)$ with redshift $z$ as a variable.

\begin{figure}[!htbp]
	\centering
	\renewcommand\figurename{Fig}
	\subfigure[Comparison of distance magnitudes. Divergences of different models are bigger than Fig.\ref{m1}.]
	{	
		\label{m2}
		\includegraphics[width=2in,height=1.3in]{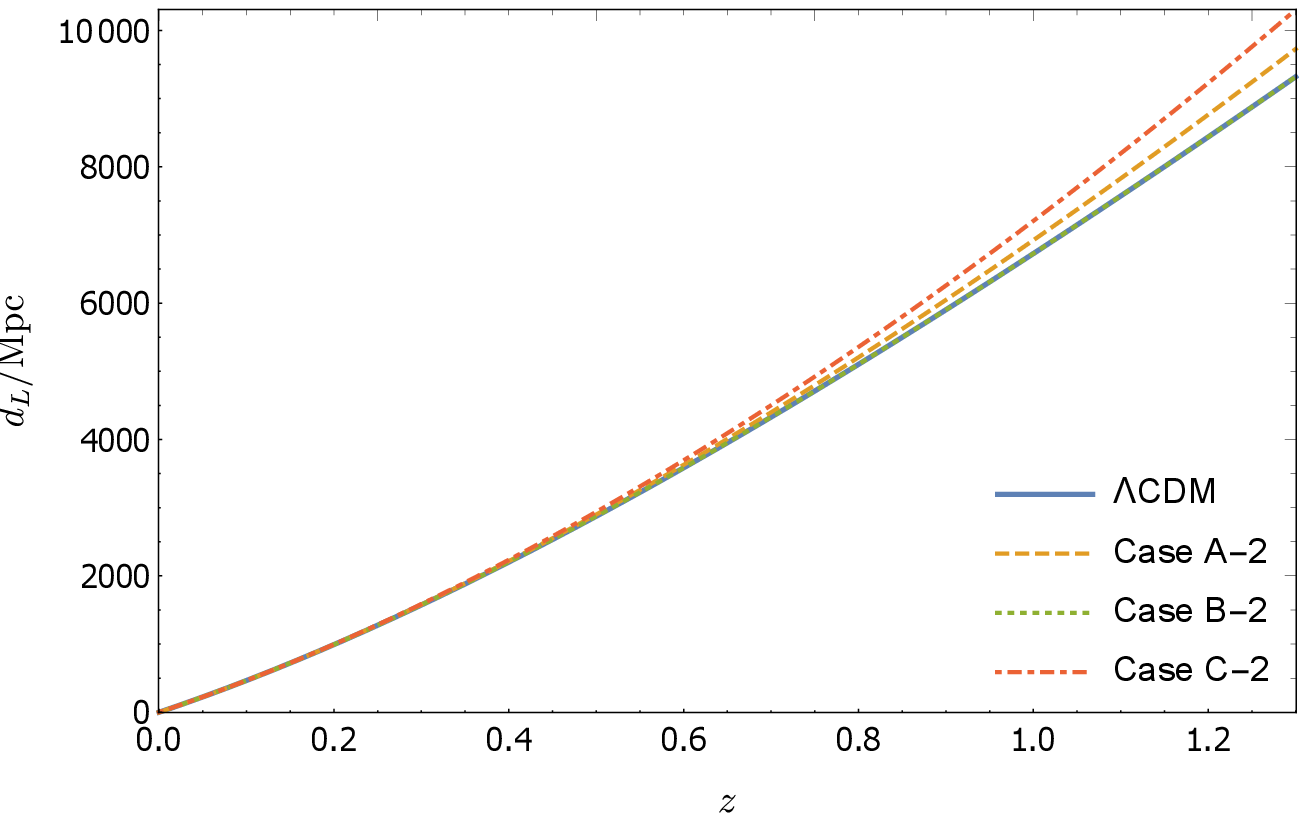}
	}
	\quad
	\subfigure[Divergences of luminosity distance are obvious. ]
	{	
		\label{dL2}
		\includegraphics[width=2in,height=1.3in]{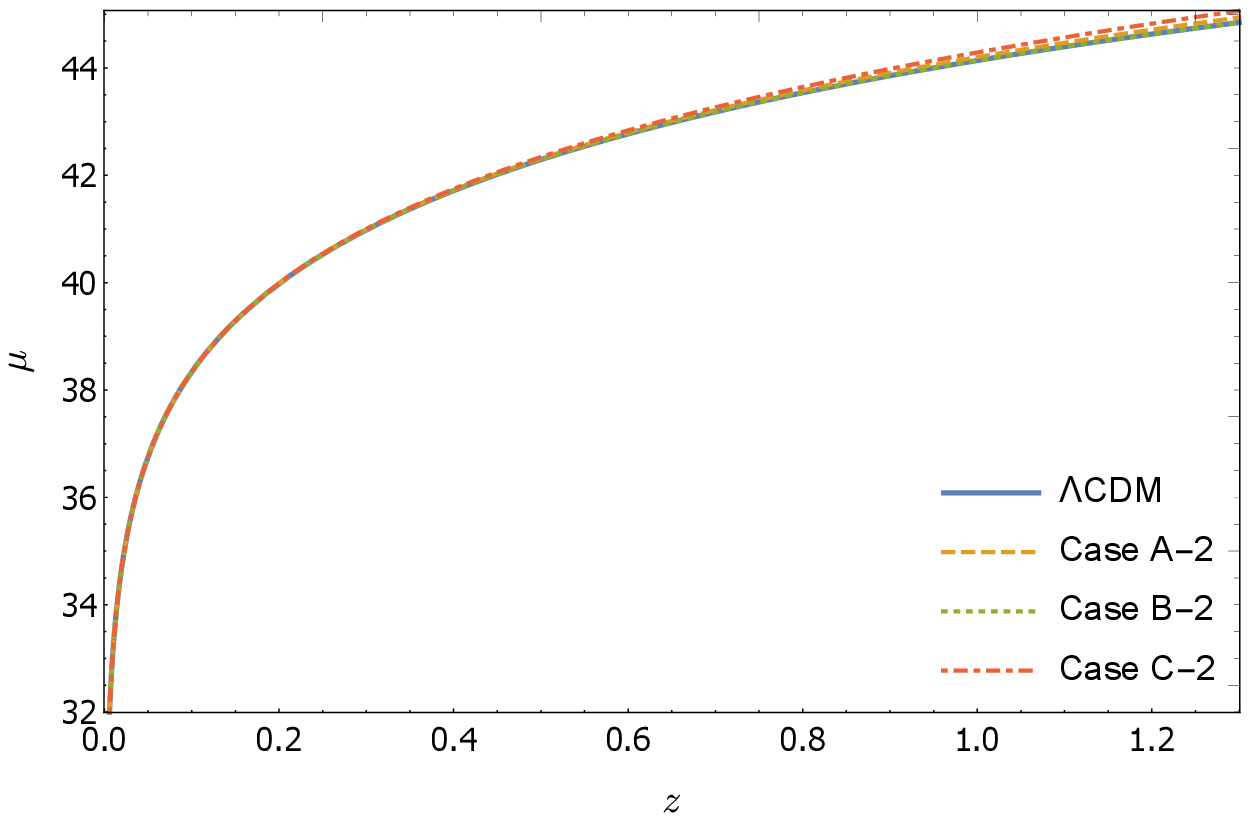}		
	}
	\caption{${\mathscr K(t_0)=-1.535H_0}$ Case}
	\label{observe2}
\end{figure}

Comparison of the results above among LSLV models and $\Lambda$CDM model as well as the measurement\cite{nielsen2016marginal} are presented in Fig.\ref{observe1} and Fig.\ref{observe2}. The distance modulus is defined as ${\mu=25+5\log_{10}\left(d_L/Mpc\right)}$(see \cite{nielsen2016marginal}). 

\section{Summary and Outlook}
Because of the vacuum energy density is 54 to 112 order higher than the effective cosmological constant, one encounter the fine tuning problem in the approach that the value of cosmological constant and one of vacuum energy density cancels almost exactly and leaving a very tiny effective cosmological constant\cite{martin2012everything}. 
Local Lorentz symmetry is the most exact symmetry of the Nature so that the LV at the macroscopic scale, if there is, is inevitably very tiny for the Planck scale suppression. Our suggestion that the LSLV at the cosmic scale induced by quantum gravity through inflation is very different from the idea seeking LV in low energy physics by standard model extension etc. Our modified $SO(3)$ gauge gravity approach is an example which is successfully in assorting the LSLV to the dark energy like effect. Since it is a common feather of LV gauge gravity that the connection deviating from Levi-Civita one, it is reasonable to assert that the LSLV impact the evolution of the universe as the dark energy does or the LSLV may be regarded as the origin of the dark energy instead of quintessence or phantom particles or some strange scalar particle which gives negative pressure.

To make sure the assertion is model independent and universal, it is necessary to investigate various modified gravity with LV and compare the corresponding dark energy effects of different models especially the comparison between teleparallel gravity framework\cite{Aldrovandi:2013wha} and curvature approach one. 

Donoghue\cite{Donoghue:2016vck} develops an idea recently that Lorentz connection may be part of the high energy gravity degree of freedom which confines or condensates to the Levi-Civita one at energy scale lower than the Planck scale similar to what happens in QCD. In this sense, the inflation may separates a phase of Lorentz connection plasma to the horizon scale almost in an instance to let the unconfined Lorentz connection phase lose interaction each other and may confined to other phase other than Levi-Civita one at large scale while confines to Levi-Civita one at short distance. It makes sense in this way our assumption on that the LV from quantum gravity can be transformed to the LSLV through inflation. Ideas from QCD can be leat to the road how to realize LSLV from QGLV and it is worthy to investigate. We give here some predictions deviating from $\Lambda$CDM model both in the observation of the early universe as well as in the late universe. There is already some clue that the dark energy is time varying which can not be explained by $\Lambda$CDM model\cite{Zhao:2017cud}. This paper put a step toward a natural explanation about it.

It is widely believed that the physics approaching Planck scale will be quantum gravity. However the evidence of ever existence of quantum gravity is not obvious. Our approach indicates that the accelerating expansion of the universe or the dark energy may be regarded as the remnant of the existence of quantum gravity.

\section*{Acknowledgment}
This work is supported by the National Natural Science Foundation of China, under Grant No. 11435005, Grant No. 11775080 and Grant No. 11865016.

\newpage
\bibliographystyle{JHEP}
\bibliography{References}

\end{document}